\documentclass[conference]{IEEEtran}

\ifCLASSINFOpdf
\else
\fi
\IEEEoverridecommandlockouts

\usepackage{amsmath}
\usepackage{amsfonts}
\usepackage{amsmath,amssymb,graphicx,bm,booktabs}
\usepackage{graphicx} 
\usepackage{epstopdf}
\usepackage{subfigure}
\usepackage{mathrsfs}
\usepackage{amsthm}

\usepackage{mathrsfs}
\usepackage{mathtools}
\usepackage{indentfirst}
\usepackage{cite}
\usepackage{acronym}
\usepackage[square, comma, sort&compress, numbers]{natbib}

\acrodef{msrs}[MSRS]{multistatic radar system}
\acrodef{rass}[RASS]{random array subset selection}

\newtheorem{proposition}{Proposition}

\IEEEoverridecommandlockouts
    \IEEEpubid{\makebox[\columnwidth]{978-1-7281-8942-0/20/\$31.00~\copyright~2020 IEEE \hfill}\hspace{\columnsep}\makebox[\columnwidth]{ }}

\begin{document}
%
\title{A Random Antenna Subset Selection Jamming\\ Method against Multistatic Radar System}

\author{\IEEEauthorblockN{Xiangtuan Wang\thanks{This work received funding from the National Natural Science Foundation of China under grants 61801258.}, Yimin Liu$^\ast$\thanks{$^\ast$Correspondence: yiminliu@tsinghua.edu.cn} and Tianyao Huang}
\IEEEauthorblockA{Department of Electronic Engineering, Tsinghua University, Beijing 100084, China}
}

%


\maketitle

\begin{abstract}
\ac{msrs} is considered an effective scheme to suppress mainlobe jamming, since it has higher spatial resolution enabling  jamming cancellation from spatial domain. To develop electronic countermeasures against \ac{msrs}, a \ac{rass} jamming method is proposed in this paper. 
In the \ac{rass} jammer,  elements of the array antenna are activated randomly, leading to stable mainlobe and random sidelobes, different from the traditional jammer that applies the complete antenna array enjoying constant mainlobe and sidelobes. 
We study the covariance matrix of jamming signals received by radars, and derive its rank, revealing that the covariance matrix is of full rank. 
We also calculate the output jamming to signal and noise ratio (JSNR) after the subspace-based jamming suppression methods used in \ac{msrs} under the proposed jamming method, which demonstrates that the full rank property invalidates such suppression methods. 
Numerical results verify our analytical deduction and exhibit the improved countermeasure performance of our proposed \ac{rass} jamming method compared to the traditional one.
\end{abstract}


\acresetall

%
\IEEEpeerreviewmaketitle

\section{Introduction}\label{sec:1}
Active electronic countermeasure (ECM) aims at reducing the performance of hostile radars by transmitting jamming signals, thus protect targets from detection by the radars. In this paper, we propose a jamming method to fight against \ac{msrs} \cite{Chernyak}.

\ac{msrs} is considered an effective scheme when encountering the threat of mainlobe jamming. Here, mainlobe jamming  means that the jamming signals propagate into radar from the radar's mainlobe, same as the radar returns from targets, and is tuned at the same frequency of the radar returns. Therefore,  this kind of jamming signals are difficult for the traditional monostatic radar to cancel, because the jammer and target are not distinguishable in the spatial domain.
To alleviate the influence of mainlobe jamming, \ac{msrs} is applied, which consists of widely distributed  radars and jointly processes signals received by all the radars, yielding a large virtual antenna aperture that enables distinguishing targets from jammer. 

\ac{msrs} relies on jamming suppression methods, as discussed in some existing literature \cite{ZHAOS, YEOK, GEM}.
The basic idea behind these methods is to use the high spatial resolution of \ac{msrs}, resulted from the large virtual aperture, to identify radar echoes from the strong jamming signals.
As long as the target and jammer are separable with respect to the virtual aperture of \ac{msrs}, the covariance matrix of radars' received signals can be divided into different subspaces, corresponding to radar returns from targets and jamming signals, respectively. Since the intensity of the jamming signal usually dominates those of radar returns and noises, the subspace of jamming signals can be found by seeking the largest eigenvalue of the covariance matrix.
For example, eigenprojection algorithm \cite{YEOK} first estimates the subspace of the jamming signals, and then projects the received signals onto its orthogonal subspace for jamming cancellation. Simulations validate the effectiveness of such method used by \ac{msrs}, thus developing new countermeasures against \ac{msrs}  becomes a raising demand.



We propose \ac{rass} for this purpose. \ac{rass}, which drives a random subset of an array antennas and changes its selection along with time,
was used in physical layer secure communication \cite{VALLIAPPANN},  resulting in a directional radiation pattern that projects a sharply defined constellation in the desired direction and expands further randomized constellation in other directions. In a joint radar communication system \cite{MAD}, the authors applied \ac{rass} to allocate antenna resources between two functions of radar and communication, and revealed that \ac{rass} generates stable  mainlobe with random sidelobes. 

Inspired by these ideas \cite{VALLIAPPANN, MAD}, we explore the application of \ac{rass} in the purpose of jamming against \ac{msrs}, where one radar resides in the mainlobe of the jammer and the rest in the sidelobes. Note that the subspace-based jamming cancellation methods used in \ac{msrs} rely on the low rank structure of the jamming signals' covariance matrix. This inherent low rank property, enabling  the orthogonal projection operation \cite{YEOK} to cancel the jamming signals while to preserve a large part of energy of the radar returns, stems from the fact that the phase differences of jamming signals received by all the radars are {\emph{fixed}} with respect to time.
Therefore, the use of \ac{rass}, which leads to randomness in its sidelobes, will introduce randomness in the phase differences, no longer being fixed. As a result, the low rank property of jamming signal's covariance matrix is destroyed, reducing the performance of the subspace-based jamming cancellation methods.

Particularly, in this paper, we derive the rank of the covariance matrix, validating the above intuitive inference. We also quantify the output jamming to signal and noise ratio (JSNR) of the eigenprojection method under the proposed jamming strategy. Both theoretical analyses and numerical results demonstrate that the \ac{rass}-based jamming pattern significantly outperforms the traditional counterpart that applies a full antenna array.

The rest of this paper is organized as follows: The signal models of \ac{msrs} and \ac{rass}-based jammer are introduced in Section \ref{sec:2}. Section \ref{sec:3} analyzes the rank of covariance matrices and the JSNR under the proposed jamming method. Numerical results are given in Section \ref{sec:4}. Section \ref{sec:5} concludes the paper.

\section{Signal model}\label{sec:2}
In this section, we introduce the geometry of the radars, jammer and target, followed by the signal models of the traditional and the proposed \ac{rass} jamming pattern in Section~\ref{sec:2.1} and \ref{sec:2.2}, respectively.

\begin{figure}[!t]
\centering
\includegraphics[width=.4\textwidth]{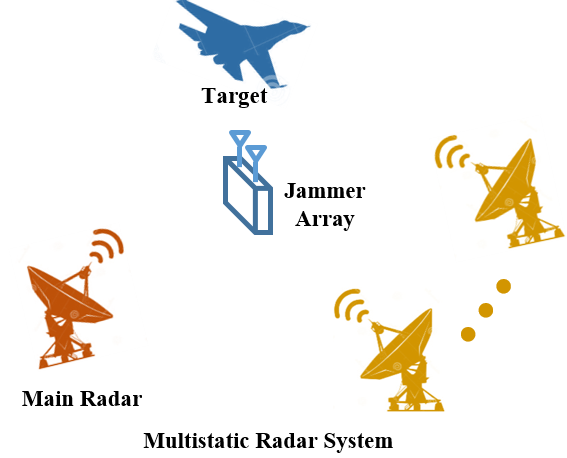}
\vspace{-0.3cm}
\caption{System model of \ac{msrs} and mainlobe jamming.}
\vspace{-0.2cm}
\label{Fig:sysmodel}

\end{figure}

Consider an \ac{msrs}, consisting of $K$ radars that are exactly synchronized. In these radars, one main radar transmits signals and receives the echoes, while the rest  only receive the target returns, working in passive mode. There is a stand-off jammer, such as escort decoy, located closely to the target, jamming towards the main radar from the radar's mainlobe to protect the target echoes from being detected.
The geometry of \ac{msrs}, target and jammer is shown in Fig.~\ref{Fig:sysmodel}.
Here, the jammer and target are closely spaced, which means that they are not distinguishable in spatial domain by a traditional monostatic radar. However, we assume that the \ac{msrs} spans a large area, synthesizing a large equivalent aperture, such that the \ac{msrs} is able to identify the signals from jammer and target in spatial domain.

In the subsequent Section~\ref{sec:2.1}, we will present the signal model under traditional jamming pattern, and a typical method used in radar to eliminate the jamming signal. To avoid being eliminated by the radar, we propose a new jamming pattern in Section~\ref{sec:2.2}.

\subsection{Traditional jamming pattern}\label{sec:2.1}
In this section, we consider the traditional jamming pattern that applies the full array to transmit jamming signal. We first present the transmit signal model of the jammer, followed by the receive signal model of the \ac{msrs} and the basic idea of cancelling the jamming.

Assume that the jammer is equipped with a uniform linear array (ULA) with $N$ elements uniformly spaced by an interval $d$. Denote by $\lambda$ the wavelength of the signal, and by $\theta_1$  the angle of the main radar with respect to the array. The jamming signal is pointed towards the main radar, hence the transmit signal of jammer is written as
\begin{equation}
\label{Eq:fullsignal}
\bm{v}(t) = \bm{\alpha}(\theta_{1})r(t),
\end{equation}
where $\bm{\alpha}(\theta):=[1,\exp^{j2\pi d\sin{\theta}/\lambda},...,\exp^{j2\pi (N-1)d\sin{\theta}/\lambda}]^T\in\mathbb{C}^{N\times 1}$ is the steering vector towards $\theta$, and $r(t)$ is the jamming signal. 
Here, we do not restrict the jamming mode of $r(t)$, which can be chosen arbitrarily, such as noise jamming, deception jamming, interrupted-sampling repeater jamming and so on \cite{FengD}.

Then, the received jamming signal of $k$-th radar can be written as
\begin{equation}
q_{k}(t-\tau^{j}_{k})=\bm{\alpha}^H(\theta_{k})\bm{\alpha}(\theta_{1})r(t-\tau^{j}_{k}),
\end{equation}
where $\theta_k$ is the angle of the $k$-th radar with respect to the jammer's array, $\tau^{j}_{k}$ denotes the delay from the jammer towards the $k$-th radar, and $^H$ denotes conjugate transposition.

The signal received by the $k$-th radar is a superposition of radar echoes reflected by the target, jamming signals and noises, given by
\begin{equation}
x_{k}(t)=s_{k}(t-\tau^{t}_{k})+q_{k}(t-\tau^{j}_{k})+n_{k}(t),
\label{Eq:signalmodel}
\end{equation}
where $s_{k}(t-\tau^{t}_{k})$ is the target echo, $n_{k}(t)$ represents receiving Gaussian noise with the variance of $\sigma^{2}$, and $\tau^{t}_{k}$  denotes the double of the delay from the target to the $k$-th radar.
For convenience, we denote by $\bm s(t) := [s_{1}(t-\tau^{t}_{1}),...,s_{K}(t-\tau^{t}_{K})]^T\in\mathbb{C}^{K\times 1}$, $\bm q(t) := [q_{1}(t-\tau^{j}_{1}),...,q_{K}(t-\tau^{j}_{K})]^T\in\mathbb{C}^{K\times 1}$ and $\bm n(t) := [n_{1}(t),...,n_{K}(t)]^T\in\mathbb{C}^{K\times 1}$ the signal, jamming and noise vectors, respectively.

Based on the received signal model above, we then introduce the eigenprojection method for jamming elimination \cite{YEOK}. Recall that these distributed receiving radars are fully synchronized and can be regarded as a virtual large aperture in receive, thus the \ac{msrs} has the potential to separate the jamming from the target echoes.

To present the  eigenprojection method, we denote by $\bm{x}(t):=[x_{1}(t),...,x_{K}(t)]^T \in \mathbb{C}^{K \times 1}$ the received signals from all radars. The covariance matrix of $\bm x(t)$ is expressed as
\begin{equation}
\bm{R}_{XX}:=\mathrm{E}[\bm{x}(t)\bm{x}^{H}(t)].
\end{equation}
Assuming that the target echoes, jamming signals and noises are mutually uncorrelated, we can get
\begin{equation}
\begin{aligned}
\bm{R}_{XX}=\bm{R}_{SS}+\bm{R}_{JJ}+\sigma^2 \bm{I},
\label{eq:rxx}
\end{aligned}
\end{equation}
where $\bm{R}_{SS}\in\mathbb{C}^{K\times K}$ and $\bm{R}_{JJ}\in\mathbb{C}^{K\times K}$ are the covariance matrices of target echoes and jamming signals, respectively,
\begin{equation}
\begin{aligned}
\bm{R}_{SS}&=\mathrm{E}[\bm{s}(t)\bm{s}(t)^{H}],\\
\bm{R}_{JJ}&=\mathrm{E}[\bm{q}(t)\bm{q}(t)^{H}].
\label{Eq:elementfullarray}
\end{aligned}
\end{equation}
We then apply eigenvalue decomposition to $\bm{R}_{XX}$, yielding
\begin{equation}
\bm{R}_{XX}=\sum_{i = 1}^{K} \lambda_i \bm{u}_{i}\bm{u}^H_i,
\end{equation}
where $\lambda_i \in \mathbb{R}$ and $\bm u_i \in \mathbb{C}^{K \times 1}$ are the $i$-th eigenvalue and eigenvector, respectively.
We assume that the energy of jamming signals dominate the target echoes, and the latter is more significant than the noise. Consequently, we can sort the eigenvalues in a descent order such that $\lambda_1\geq \dots \geq \lambda_J \gg \lambda_{J+1} \geq \dots\geq\lambda_{T+J} \gg \lambda_{T+J+1}\geq \dots \geq \lambda_K $, where $J$ and $T$ denote the number of jammers and targets. Here, we set $J= T = 1$. By exchanging the eigenvectors correspondingly,  we divide $\bm{R}_{XX}$ into three subspaces, given by
\begin{equation}
\bm{R}_{XX}=\bm{U}_{S}\bm{\Lambda}_{S}\bm{U}_{S}^{H}+\bm{U}_{J}\bm{\Lambda}_{J}\bm{U}_{J}^{H}+\bm{U}_{N}\bm{\Lambda}_{N}\bm{U}_{N}^{H},
\label{Eq:normalcov}
\end{equation}
where $\Lambda_J := {\rm diag}(\lambda_{1},\dots,\lambda_{J}) \in \mathbb{R}^{J \times J}$, $\Lambda_S := {\rm diag}(\lambda_{J+1},\dots,\lambda_{T+J}) \in \mathbb{R}^{T \times T}$, and $\Lambda_N := {\rm diag}(\lambda_{T+J+1},\dots,\lambda_{K}) \in \mathbb{R}^{(K-J-T) \times (K-J-T)}$ denote eigenvalues corresponding to jamming, echoes and noise signals, respectively. The jamming, signal and noise subspaces are given by $\bm{U}_{J}:= \left[ \bm u_{1}, \dots, \bm u_{J} \right] \in \mathbb{C}^{K\times J}$,  $\bm{U}_{S}:= \left[ \bm u_{J+1}, \dots, \bm u_{T+J} \right] \in \mathbb{C}^{K \times T}$ and $\bm{U}_{N}:= \left[ \bm u_{T+J+1}, \dots, \bm u_{K} \right] \in \mathbb{C}^{ K\times (K-J-T)}$, respectively. 

Based on the separated jamming subspace, the eigenprojection method cancels the jamming signals using orthogonal projection, yielding
\begin{equation}
\bm{y}:=\bm{P}^{\perp}\bm{x}:=(\bm{I}-\bm{U}_{J}\bm{U}_{J}^{H})\bm{x},
\label{Eq:normaleigenprojection}
\end{equation}
where $\bm y$ denotes the residual signal after jamming elimination, $\bm P:=\bm{U}_{J}\bm{U}_{J}^{H} \in \mathbb{C}^{K \times K}$ denotes the projection matrix with respective to $\bm U_J$, and $\bm P^{\perp} := \bm I - \bm P$ represents the orthogonal projection matrix.

From the jamming elimination procedure above, we find that the identification of jamming subspace plays a key role. The identification task is possible for \ac{msrs}, because of two features of  the jammer: One feature is that the amplitude of the jamming signal is most significant in comparison with radar echoes and noises; The other stems from the fact that the received jamming signals by different radars are `coherent', which means that the phase differences between theses received signals are \emph{fixed}.
In order to fight against such distributed radars, we need to propose new jamming strategies by changing these features.
While maintaining the dominance of the jamming signal over radar echoes are generally necessary, we use \ac{rass} to destroy the `coherence', resulting in difficulties for the \ac{msrs} to find the jamming subspace correctly, as will be introduced in the sequel.

\subsection{Proposed jamming pattern}\label{sec:2.2}
In this subsection, we introduce the \ac{rass} jamming pattern that randomly selects a subset of array antenna  to transmit jamming signal.
As \ac{rass} is an effective method that forms the mainlobe in the desired direction and keeps the randomness in undesired directions, the \ac{rass} method is used in secure communication to achieve security and directional transmission \cite{VALLIAPPANN}. Inspired by the above idea, we apply the \ac{rass} into jammer's array antenna to enable efficient mainlobe jamming and to achieve random sidelobe which benefits fighting against \ac{msrs}.
The randomness in the sidelobe destroy the `coherence' of the received signals between different static radars in \ac{msrs}, thus reducing the jamming cancellation performance of \ac{msrs}.
To see this, we first present the \ac{rass} transmit signal model with the random switch vector. Then we analyze the receive signal of  \ac{msrs} to show that the jammer achieves random beam pattern by \ac{rass} method.

\begin{figure}[!t]
\centering

\subfigure[]{
\includegraphics[width=.08\textwidth]{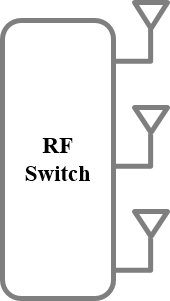}
\label{Fig:rassswitch}
}
\subfigure[]{
\includegraphics[width=.36\textwidth]{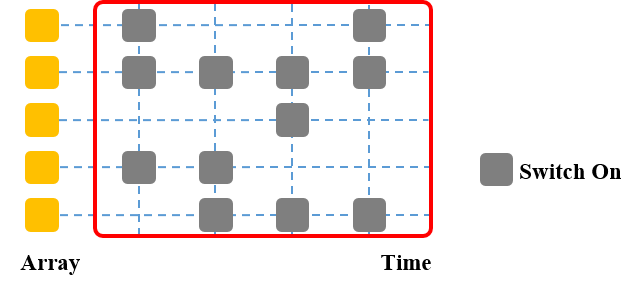}
\label{Fig:rasstimedomain}

}
\caption{The diagram of RASS method. (a)High speed RF switch is used to realize \ac{rass}. (b)The diagram of \ac{rass} jamming method in time domain.}
\end{figure}

In \ac{rass} jamming method,  $M$ out of $N$ antenna elements are selected to transmit jamming signal, $N>M$. And the selection is changed from a time slot to another. To realize the random selection, a high speed RF switch is usually required, as shown in Fig.~\ref{Fig:rassswitch}. The resultant diagram of \ac{rass} method in time domain is given in Fig.~\ref{Fig:rasstimedomain}.

We use $\bm{p}(t)\in \{0,1\}^{N\times 1}$ to denote the random switch vector, and $p_{n}(t)$ to denote its $n$th element. When the $n$th antenna is active $p_{n}(t)=1$, and 0 otherwise.
The transmitted signal of jammer array can be written as
\begin{equation}
\bar{\bm{v}}(t)=\bm{p}(t)\circ\bm{\alpha}(\theta_1)r(t),
\label{eq:transmitrass}
\end{equation}
where  the notation $\bar{\cdot}$ is used to differentiate from the counterpart of regular jamming pattern, and $\circ$ represents the Hadamard product. Compared with (\ref{Eq:fullsignal}), the transmitted signal \eqref{eq:transmitrass} of \ac{rass} jammer contains modulation by the random switch vector $\bm{p}(t)$.

We assume that  $p_{n}(t)$ follows a Bernoulli distribution with probability $p$, whose probability density function (PDF) is given by
\begin{equation}
\label{Eq:Bernoulli}
f(p_n(t)|p)=\left\{
           \begin{array}{ll}
             p^{p_n(t)}(1-p)^{1-p_n(t)}, & p_n(t)=0,1, \\
             0, & \mathrm{else.}
           \end{array}
         \right.
\end{equation}
We also assume that $p_{n}(t)$ is independent with respect to the element index $n$ and time $t$. 

Then, the jamming signal received by the $k$th radar is written as
\begin{equation}
\begin{aligned}
\bar{q}_{k}(t-\tau^{j}_{k}):=\bm{\alpha}^{H}(\theta_k)(\bm{p}(t-\tau^{j}_{k})\circ\bm{\alpha}(\theta_1))r(t-\tau^{j}_{k}).
\end{aligned}
\label{eq:receive rass}
\end{equation}
Here,  we define $\bar{\bm q}:=[\bar{q}(t-\tau^{j}_{1}),...,\bar{q}(t-\tau^{j}_{K})]^T\in\mathbb{C}^{K\times 1}$.
Substituting the steering vector, given below \eqref{Eq:fullsignal}, into \eqref{eq:receive rass} yields
\begin{equation}
\begin{aligned}
\bar{q}_{k}(t-\tau^{j}_{k})=&\sum_{n=1}^{N}p_{n}(t-\tau^{j}_{k})r(t-\tau^{j}_{k})\\
&\quad \cdot \exp^{j2\pi (n-1)d(\sin{\theta_k}-\sin{\theta_1})/\lambda}.
\label{Eq:jammingsignal}
\end{aligned}
\end{equation}

Similar with \eqref{Eq:signalmodel}, the received signal of the $k$-th radar under RASS jamming method is
\begin{equation}
\bar{x}_{k}(t)=s_{k}(t-\tau^{t}_{k})+\bar{q}_{k}(t-\tau^{j}_{k})+n_{k}(t).
\label{Eq:signalmodelRASS}
\end{equation}
We stack signals from all the radars, which yields the vector ${\bar{\bm{x}}}(t):=[{\bar x}_{1}(t),...,{\bar x}_{K}(t)]^T \in \mathbb{C}^{K \times 1}$.
Here, we also assume that echoes from target $s_{k}(t-\tau^{t}_{k})$, noises $n_{k}(t)$ and the proposed jamming signals $\bar{x}_{k}(t)$ are mutually independent.

To interpret (\ref{Eq:jammingsignal}), we consider two cases: $k = 1$ and $k \neq 1$. The former represents the main radar, where the jamming beam is directed, while the latter represents the rest radars, located in the sidelobe of the jammer's beam pattern.
When $k=1$,
the coefficient $\bm{\alpha}^{H}(\theta_k)(\bm{p}(t-\tau^{j}_{k})\circ\bm{\alpha}(\theta_1))$ reduces to $\bm{\alpha}^{H}(\theta_1)(\bm{p}(t-\tau^{j}_{1})\circ\bm{\alpha}(\theta_1))$, which equals to $\bm 1^T \bm{p}(t-\tau^{j}_{1})$, yielding the number of selected antenna units at instance $t-\tau^{j}_{1}$. This is a consequence of the fact that the jammer steers the beam towards the main radar, and also indicates that the phase of the coefficient stays unchanged even if  $p_n(t)$ is randomly changed with respect to $t$. As a result, the antenna gain of the \ac{rass} jammer against the main radar is guaranteed as long as enough antenna units are utilized.
In the case of $k\neq 1$, the coefficient $\bm{\alpha}^{H}(\theta_k)(\bm{p}(t-\tau^{j}_{k})\circ\bm{\alpha}(\theta_1))$ changes randomly with respect to $t$, which means that the $K-1$ passive radars will receive a randomized  jamming signal.
Therefore, the phase differences between jamming signals received by different radars change with respect to $t$, destroying the `coherence' between radars. This property degrades the jamming elimination method used by   \ac{msrs}, thus better protects the target from being detected by radars.

In the next section, we will discuss the anti-suppression performance of the proposed RASS jamming method in the term of  jamming cancellation method presented in Section~\ref{sec:2.1}.

\section{Performance Analysis of the Proposed Jamming Strategy}\label{sec:3}
In this section, we
analyze the performance of the \ac{rass}-based jamming strategy against \ac{msrs} equipped with the eigenprojection method for jamming cancellation, as discussed in Section~\ref{sec:2.1}.
To this aim, we first
derive the covariance matrix associated with \ac{rass}  method in Section~\ref{sec:3.1}.

\subsection{Covariance matrix}\label{sec:3.1}
We derive the covariance matrix of the received signals, i.e., $\bm{\bar{R}}_{XX}:=\mathrm{E}[\bm{\bar{x}}(t)\bm{\bar{x}}^{H}(t)]$, as
\begin{equation}
\begin{aligned}
\bm{\bar{R}}_{XX}&=\bm{R}_{SS}+\bm{\bar{R}}_{JJ}+\sigma^2 \bm{I},
\label{Eq:rasscov}
\end{aligned}
\end{equation}
following the assumption that target, jamming and noise signals are uncorrelated.
Here, similarly with \eqref{Eq:elementfullarray}, we use $\bm{\bar{R}}_{JJ}$ to denote the covariance matrix of the received jamming signals $\bar{\bm{q}}(t)$, given by
\begin{equation}
\begin{aligned}
&\bm{\bar{R}}_{JJ}:=\mathrm{E}[\bar{\bm{q}}(t)\bar{\bm{q}}^{H}(t)].
\label{Eq:elementrass}
\end{aligned}
\end{equation}

Comparing (\ref{Eq:rasscov}) with (\ref{Eq:normalcov}), we find that only the covariance matrix of jamming signals $\bm{\bar{R}}_{JJ}$ is changed due to the introduction of \ac{rass}. 
Under the assumption that $\bm{p}(t)$ obeys Bernoulli distribution, we derive $\bm{\bar{R}}_{JJ}$ as stated in the following proposition:

\begin{proposition}\label{pro:1}
The covariance matrix of jamming signal $\bm{\bar{R}}_{JJ}$ can be written as
\begin{equation}
\label{Eq:covBernoulli}
\bm{\bar{R}}_{JJ}=p^2 \bm{R}_{JJ}+Np(1-p)R_{rr}\bm{I},
\end{equation}
where $R_{rr}$ is the autocorrelation of stationary jamming signal $r(t)$, given by
\begin{equation}
R_{rr}:=\mathrm{E}[r(t)r^{*}(t)].
\end{equation}
\end{proposition}
\begin{proof}
Due to the length limit of this paper, we leave the proof details in the journal version of this paper.
\end{proof}

From \eqref{Eq:covBernoulli}, we find that $\bm{\bar{R}}_{JJ}$ is of full rank as long as $Np(1-p)R_{rr}\bm{I} \neq \bm 0$. Since $R_{rr} > 0$, the term equals 0 if and only if $p = 0$ or $1$, which means that no antenna or all the antennas are operating, respectively.  In contrast to the traditional jamming pattern, i.e., when $p = 1$, where the rank of $\bm{{R}}_{JJ}$ is typically 1 under the assumption that there exists only 1 jammer operating, the proposed RASS jamming method increases the rank of covariance matrix.  As a result, the typical eigenprojection method used for jamming cancellation in \ac{msrs}, which relies on identifying the dominant eigenvalues corresponding to jamming signals, will be affected severely.

To see this, we  further discuss the eigenvalues and eigenvectors of $\bm{\bar{R}}_{XX}$ in Section \ref{sec:3.2}.

\subsection{Eigenvalues and eigenvectors of the covariance matrix}\label{sec:3.2}
In this subsection, we apply matrix perturbation method \cite{CHAMPAGNEB} to study eigenvalues and eigenvectors of $\bm{\bar{R}}_{XX}$ in light of the counterparts of $\bm{{R}}_{XX}$.

We substitute  (\ref{Eq:covBernoulli}) into (\ref{Eq:rasscov}), and compare the result with \eqref{eq:rxx}, implying
\begin{equation}
\begin{aligned}
\bm{\bar{R}}_{XX}&=\bm{R}_{XX}+(p^2-1)\bm{R}_{JJ}+Np(1-p)R_{rr}\bm{I}.
\end{aligned}
\end{equation}
Here, we regard $\bm{\bar{R}}_{XX}$ as a matrix generated from $\bm{{R}}_{XX}$ by adding a perturbation matrix, defined as $\Delta\bm{R}_{XX}:=(p^2-1)\bm{R}_{JJ}+Np(1-p)R_{rr}\bm{I}$. This facilitates revealing the relationship between eigenvalues and eigenvectors of these two matrices.

Let $\bar{\lambda}_{k}$ and $\bm{\bar{u}}_{k}$ denote the $k$-th eigenvalue and the corresponding eigenvector  of $\bm{\bar{R}}_{XX}$, respectively. Here, we define $\Delta\lambda_k$ and $\Delta\bm{u}_{k}$ such that  $\bar{\lambda}_{k}=\lambda_{k}+\Delta\lambda_k$ and $\bm{\bar{u}}_{k}=\bm{u}_{k}+\Delta\bm{u}_{k}$.
Through the definitions of eigenvalues, we have
\begin{equation}
\label{Eq:eigenfunction0}
\bm{\bar{R}}_{XX}\bm{\bar{u}}_{k}=\bar{\lambda}_{k}\bm{\bar{u}}_{k},
\end{equation}
or equivalently,
\begin{equation}
\begin{aligned}
\label{Eq:eigenfunction}
(\bm{R}_{XX}+\Delta\bm{R}_{XX})(\bm{u}_{k}+\Delta\bm{u}_{k})&=(\lambda_{k}+\Delta\lambda_k)(\bm{u}_{k}+\Delta\bm{u}_{k}).
\end{aligned}
\end{equation}
When the second order terms, i.e., $\Delta\bm{R}_{XX}\Delta\bm{u}_{k}$ and $\Delta\lambda_k\Delta\bm{u}_{k}$, are negligible, the perturbations in eigenvalues and eigenvectors can be approximated as \cite{CHAMPAGNEB}
\begin{equation}
\begin{aligned}
\Delta\lambda_{k}&\approx\bm{u}^{H}_{k}\Delta\bm{R}_{XX}\bm{u}_{k},\\
\Delta\bm{u}_{k}&\approx\bm{U}\bm{b}_{k},
\label{Eq:deltauk}
\end{aligned}
\end{equation}
where $\bm{U}:=[\bm{u}_{1},...,\bm{u}_{K}]\in\mathbb{C}^{K\times K}$ is the eigen matrix of $\bm{R}_{XX}$ and $\bm{b}_{k}\in \mathbb{C}^{K\times 1}$ has the $i$-the entry given by $\frac{1}{\lambda_{i}-\lambda_{k}}\bm{u}^{H}_{i}\Delta\bm{R}_{XX}\bm{u}_{k}$, $i \neq k$, and the $k$-th entry being zero.
Due to the length limit of this paper, we leave the proof for \eqref{Eq:deltauk} in the journal version of this paper.
Then, the $k$-th eigenvector of $\bm{\bar{R}}_{XX}$ is given by
\begin{equation}
\bm{\bar{u}}_{k}=\bm{u}_{k}+\bm{U}\bm{b}_{k},
\label{Eq:vectornew}
\end{equation}
benefitting the analysis on JSNR, which we show in the sequel.

\subsection{Improvement of JSNR}\label{sec:3.3}

Under the proposed jamming pattern, we assume that \ac{msrs} applies the same jamming cancellation strategy: Use the eigenvector $\bm{\bar{u}}_{1}$ corresponding to the largest eigenvalue $\bar{\lambda}_1$ to realize eigenprojection.

Analogy to \eqref{Eq:normaleigenprojection}, the output signal is given by
\begin{equation}
\bm{\bar{y}}(t)=\bm{\bar{P}}^{\perp}\bm{\bar{x}}(t)=\left(\bm{I}-\bm{\bar{u}}_{1}\bm{\bar{u}}_{1}^{H}\right)\bm{\bar{x}}(t).
\label{Eq:eigenprojection}
\end{equation}

We are now ready to evaluate the performance of the eigenprojection method.  Here, we use the output JSNR as the metric, and we compare the output JSNRs under the traditional and \ac{rass} jamming patterns.
Particularly, we define the output JSNR under \ac{rass} jamming as the following: 
\begin{equation}
\begin{aligned}
\Omega_{R}&=\frac{\mathrm{E}[||\bm{\bar{P}}^{\perp}\bar{\bm{q}}||_{2}^{2}]}{E[||\bm{\bar{P}}^{\perp}(\bm{s}+\bm{n})||_{2}^{2}]},
\label{Eq:JSR}
\end{aligned}
\end{equation}
where in the numerator, the expectation is taken over $\bm{\bar{q}}$, since it is randomly changed along with time slots, while the expectation in the denominator is taken over the noise. We use subscript $_R$ to denote the \ac{rass} jamming pattern. To differentiate, we denote by $\Omega_F$ the counterpart under the traditional jamming pattern where the full antenna array is used. As the \ac{msrs} can effectively eliminate jamming signal by eigenprojection method, the $\Omega_{F} = 0$, which will be discussed later. Thus $\Omega_{R}$ represents the improvement of the proposed \ac{rass} method over the traditional counterpart in the term of JSNR.

Under assumptions on the jamming pattern, we derive its JSNR in the following proposition.
\begin{proposition}\label{pro:2} The output JSNR is given by
\begin{equation}
\begin{aligned}
\mathrm{\Omega}_{R}&=\frac{KR_{rr}}{\mathrm{E}[||\bm{s}||_{2}^{2}+||\bm{n}||_{2}^{2}]}Np(1-p).
\label{Eq:JSRexceptation}
\end{aligned}
\end{equation}
\end{proposition}
\begin{proof}
Due  to  the  length  limit  of  this  paper,  we  leave  the  proof in the journal version of this paper.
\end{proof}
\noindent The JSNR  $\Omega_R>0$ unless $p=0$ or $1$, where the random  antenna array reduces to an empty or a full array, indicating $\Omega_F = 0$.
Comparing between $\Omega_R$ and $\Omega_F$ indicates that \ac{rass} jamming pattern leads to  residual jamming energy after the jamming cancellation step via eigenprojection.
This happens because the covariance matrix $\bar{\bm{R}}_{JJ}$ is of full rank, as given in Proposition \ref{pro:1}. Therefore, the eigenprojection method, which selects a rank-one matrix for jamming cancellation, eliminate only a portion of the jamming energy, leading to dominant  residual of jamming signals that could still mask the target echoes.
This property demonstrates the improvement of proposed jamming strategy over the traditional one: The residual jamming energy after the eigenprojection procedure is significantly larger than that of the latter, although the power of transmitted jamming signal in \ac{rass} jammer is lower due to the use of only a subset of antennas,  verifying  the importance of spatial agility as introduced in \ac{rass} in the case of jamming against \ac{msrs}.
From (\ref{Eq:JSRexceptation}), we observe that the JSNR takes its maximum when $p=0.5$, suggesting the optimal jamming strategy of JSNR under Bernoulli distribution.
It is also found  that the JSNR increases with respect to the number of jammer's array elements, $N$. The relationship between JSNR and the number of radars, $K$, in the numerator, is not distinct, because the energy of target echoes and noise expressed in the denominator also increase as $K$ becomes larger.

Numerical results are given in Section \ref{sec:4} to validate the above analysis.

\section{Numerical Results}\label{sec:4}
In this section, the performance of the traditional and the proposed RASS jamming methods is demonstrated by numerical experiments. We use  range profiles obtained by \ac{msrs} and the output JNSR to evaluate the effectiveness of the jamming methods.

We consider an \ac{msrs} with $K = 4$ radars. Their three-dimensional coordinates are setting as follows:  $(x_1,y_1,z_1)=(0,0,0)$, $(x_2,y_2,z_2)=(10,0,0)$ km,  $(x_3,y_3,z_3)=(0,10,0)$ km and $(x_4,y_4,z_4)=(10,10,0)$ km. The target and jammer are closely located at $(2,3,15.3)$ km and $(2,3,15)$ km, respectively. The jammer implements a linear uniform array antenna with $N = 16$ elements and the interval $d = 0.03$ m.
We regard the first radar as the main radar, transmitting linear frequency modulation (LFM) signal with the bandwidth of 10MHz, the duration of 10$\mu$s and the center frequency of 5GHz. We uniformly divide the duration of LFM into $L = 128$ time slots. The signal-to-noise ratio (SNR) of echoes from target, defined as $\frac{||\bm{s}||^2_{2}}{\mathrm{E}[||\bm{n}||^2_{2}]}$, is set as 20dB. The $\sigma^{2}=10^{-2}$ is the noise variance. 
The jammer use Gaussian noise of the same bandwidth as its baseband signal $r(t)$, continuously jamming towards the main radar. In \ac{rass} jamming pattern, the jammer randomly selects a subarray antenna in each time slot. Recall that each antenna element has the probability $p$ to be used. The input JSNR with respect to each element of jammer's array antenna is set as $\frac{KR_{rr}}{\mathrm{E}[||\bm{s}||_{2}^{2}+||\bm{n}||_{2}^{2}]}=31$ dB. 

\begin{figure}[!t]
\centering
\includegraphics[width=.4\textwidth]{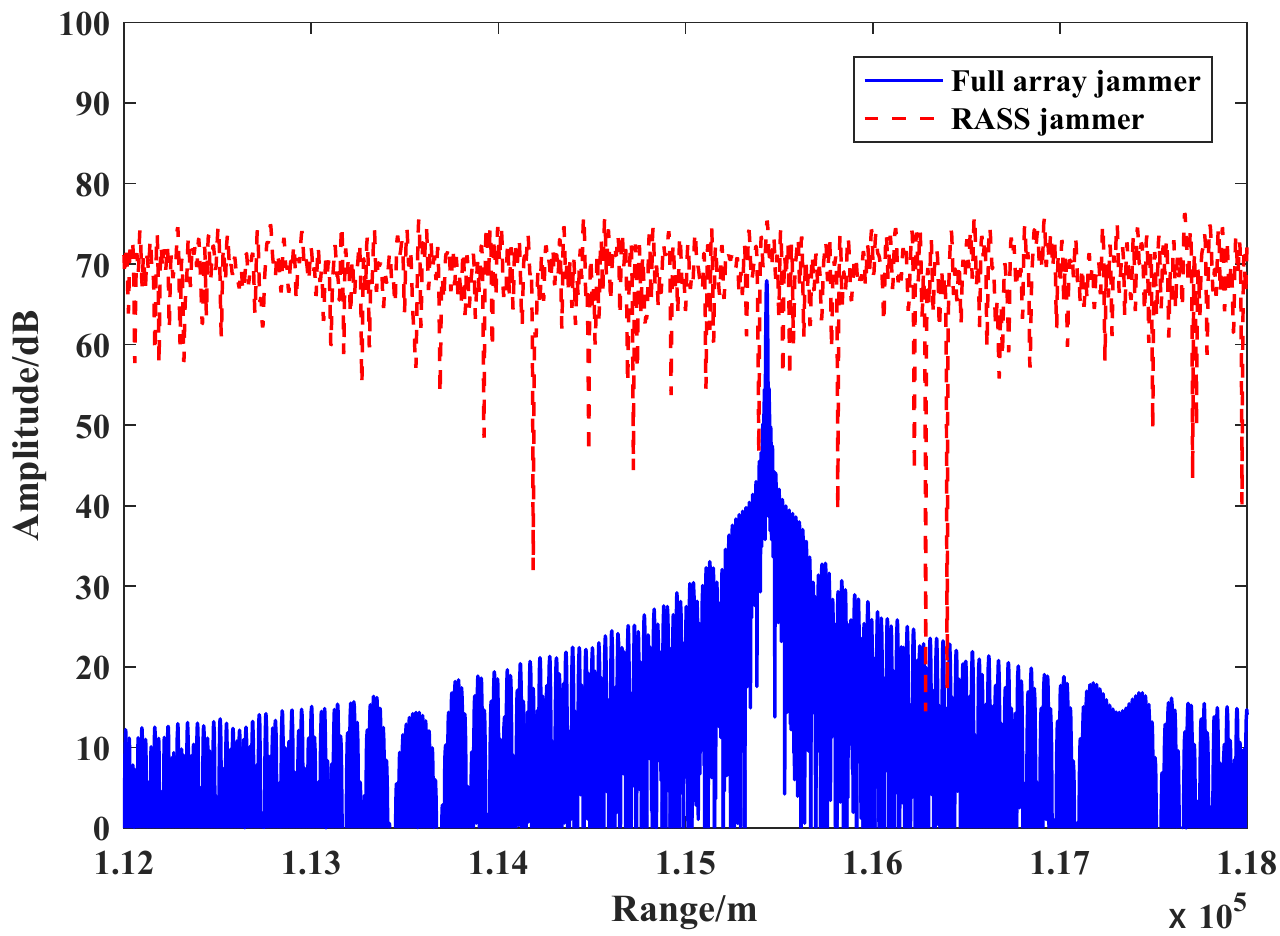}
\caption{Range profiles after eigenprojection suppression method.}
\label{Fig:rangeprofile}
\end{figure}

In the first experiment, we evaluate the obtained range profiles of targets after jamming suppression under the traditional and \ac{rass} jamming methods. We set $p = 0.5$ in \ac{rass}. Amplitudes of range profiles are shown in Fig.~\ref{Fig:rangeprofile}. In Fig.~\ref{Fig:rangeprofile}, the blue curve, representing the result by the jammer with full array antenna, gives a focused range profile. The peak of range profile indicates the target location, which means the jamming signals are cancelled successfully by \ac{msrs}.
However, the red dotted curves, obtained by the proposed jamming method, provides a noise-like range profile masking the target, indicating the success  of the jammer against \ac{msrs}.
This is because \ac{rass} introduces randomness in its sidelobe which is directed towards the auxiliary passive radars, destroying the coherence between radars. As a result, the \ac{msrs} fails to identify the eigenvalue corresponding to jamming signals, and the jamming residual of eigenprojection method is strong enough to mask the returns from target.

\begin{figure}[!t]
\centering
\includegraphics[width=.4\textwidth]{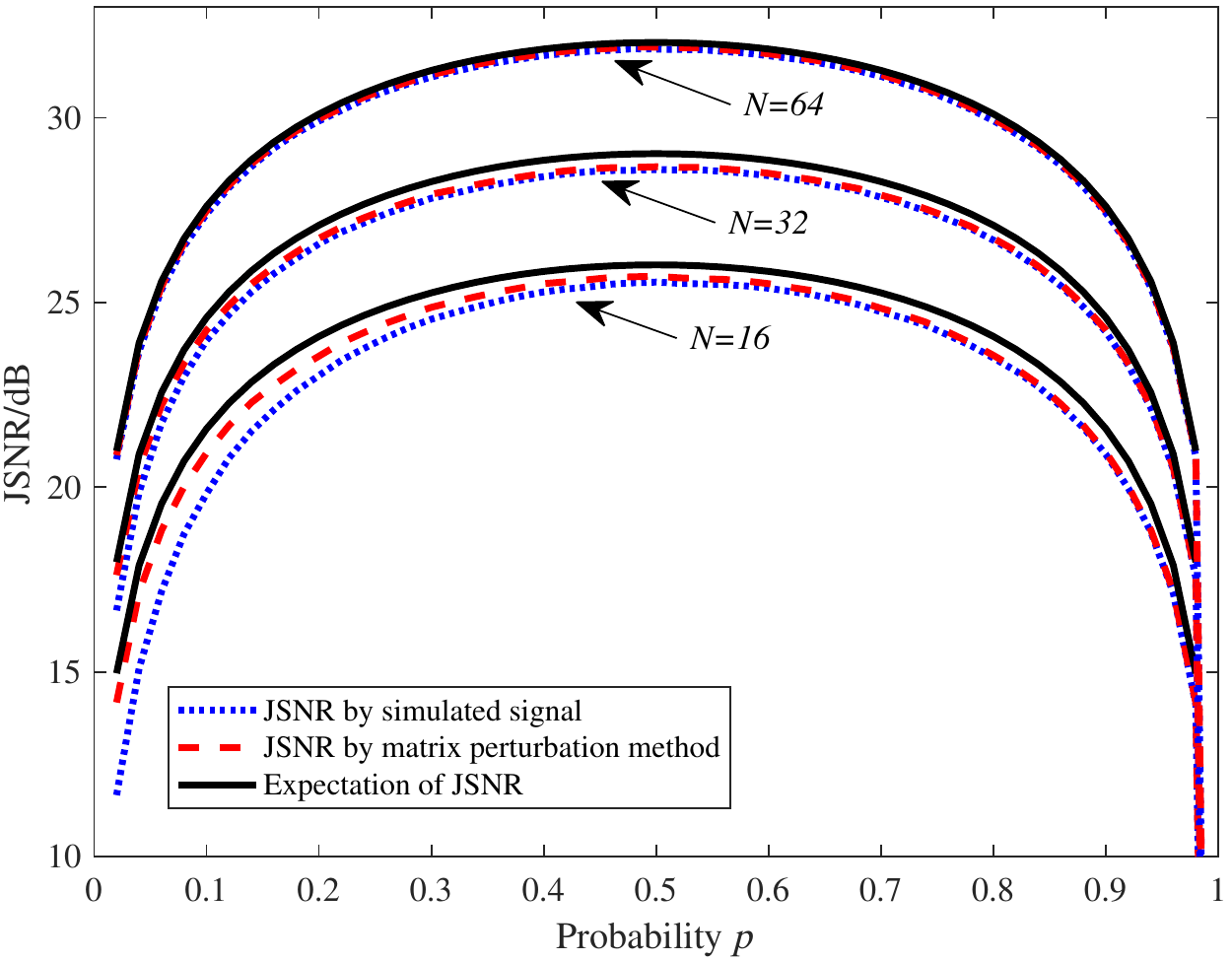}
\vspace{-0.2cm}
\caption{The JSNR results vary with probability $p$ and jammer array elements $N$.}
\label{Fig:JSR}
\vspace{-0.6cm}
\end{figure}

The second experiment is designed to calculate the output JSNR $\Omega_R$ versus $p$.  We compare the theoretical result \eqref{Eq:JSRexceptation}, denoted `Expectation of JSNR', with simulated ones obtained from \eqref{Eq:JSR}, where the expectation operation is discarded and the resulting $\Omega_R$ are averaged over 1000 Monte Carlo trials.
In particular, to validate the correctness of \eqref{Eq:vectornew}, which approximates the eigenvector with matrix perturbation method, we use two methods to construct the projection matrices in \eqref{Eq:JSR}. One is obtained from  \eqref{Eq:vectornew},  denoted by `JSNR by matrix perturbation method'. The other is achieved by performing eigendecomposition over the sampled covariance matrix $\frac{1}{L}\sum_{0}^{L-1}\bar{\bm{x}}[l]\bar{\bm{x}}^{H}[l]$, where $\bar{\bm{x}}[l]$ is $\bar{\bm{x}}(t)$ sampled at the $l$-th time slot. This result is denoted by `JSNR by simulated signal'.
All these results are shown in Fig.~\ref{Fig:JSR}, where the $y$-axis  ranges from $10$ dB to $33$ dB to clearly show the curves. Therefore, the results of $p=0 $ and $1$, which take much lower values, are not included. From the figure, we find that all these curves are close, validating the correctness of both \eqref{Eq:vectornew} and \eqref{Eq:JSRexceptation}.
As expected, the JSNR takes its maximum at $p = 0.5$ and minimum at $p = 0$ and $1$.
When $p=1$, representing the traditional jamming method that uses the full array antenna, the JSNR becomes $\Omega_{F} = -24.19 $ dB, calculated by simulated methods. Compared with the traditional jamming pattern, the \ac{rass} jamming pattern reaches much higher JSNR with lower transmitting power, validating  the importance
of spatial agility to jam against \ac{msrs}.

We then discuss the JSNR with respect to $N$, the number of jammer' antenna elements.
We set $N=16, 32$ and $64$, with an interval of $3$ dB ($10\log_{10}2$) between successive values, and the other settings are the same with the previous experiments. Similar to the previous experiment, we use  \eqref{Eq:vectornew} (with 1000 Monte Carlo trials) and \eqref{Eq:JSRexceptation} to calculate the JSNR, shown in Fig.~\ref{Fig:JSR}. From this figure, we observe $3$ dB difference in JSNRs between the successive $N$ under test,  coinciding with \eqref{Eq:JSRexceptation}.

\section{Conclusion}\label{sec:5}
In this paper, the RASS jamming method,  which randomly chooses different subset of  array antenna to transmit jamming signal at different time instances, was proposed against \ac{msrs} equipped with eigenprojection for jamming cancellation.
The \ac{rass} method introduces agile beam patterns in the sidelobe, leading to a full rank covariance matrix of jamming signals,
which significantly reduces the performance of the jamming suppression method.
Particularly, we used perturbation matrix method to analyze the eigenvalues of the signals received by radar, and quantified the output JSNR under this jamming method.
Simulation results validates our analyses and demonstrate that the proposed method significantly improves the jamming performance over the traditional counterpart that uses the complete antenna array.
\end{document}